\input harvmac.tex



\def\unlockat{\catcode`\@=11}
\def\lockat{\catcode`\@=12}

\unlockat

\def\newsec#1{\global\advance\secno by1\message{(\the\secno. #1)}
\global\subsecno=0\global\subsubsecno=0\eqnres@t\noindent
{\bf\the\secno. #1}
\writetoca{{\secsym} {#1}}\par\nobreak\medskip\nobreak}
\global\newcount\subsecno \global\subsecno=0
\def\subsec#1{\global\advance\subsecno by1\message{(\secsym\the\subsecno. #1)}
\ifnum\lastpenalty>9000\else\bigbreak\fi\global\subsubsecno=0
\noindent{\it\secsym\the\subsecno. #1}
\writetoca{\string\quad {\secsym\the\subsecno.} {#1}}
\par\nobreak\medskip\nobreak}
\global\newcount\subsubsecno \global\subsubsecno=0
\def\subsubsec#1{\global\advance\subsubsecno by1
\message{(\secsym\the\subsecno.\the\subsubsecno. #1)}
\ifnum\lastpenalty>9000\else\bigbreak\fi
\noindent\quad{\secsym\the\subsecno.\the\subsubsecno.}{#1}
\writetoca{\string\qquad{\secsym\the\subsecno.\the\subsubsecno.}{#1}}
\par\nobreak\medskip\nobreak}

\def\subsubseclab#1{\DefWarn#1\xdef #1{\noexpand\hyperref{}{subsubsection}%
{\secsym\the\subsecno.\the\subsubsecno}%
{\secsym\the\subsecno.\the\subsubsecno}}%
\writedef{#1\leftbracket#1}\wrlabeL{#1=#1}}
\lockat

\def\IL{\relax{\rm I\kern-.18em L}}
\def\IH{\relax{\rm I\kern-.18em H}}
\def\IR{\relax{\rm I\kern-.18em R}}
\def\IC{\relax\hbox{$\inbar\kern-.3em{\rm C}$}}
\def\IZ{\relax\ifmmode\mathchoice
{\hbox{\cmss Z\kern-.4em Z}}{\hbox{\cmss Z\kern-.4em Z}}
{\lower.9pt\hbox{\cmsss Z\kern-.4em Z}}
{\lower1.2pt\hbox{\cmsss Z\kern-.4em Z}}\else{\cmss Z\kern-.4em Z}\fi}


\font\manual=manfnt \def\dbend{\lower3.5pt\hbox{\manual\char127}}

\def\IZ{\relax\ifmmode\mathchoice
{\hbox{\cmss Z\kern-.4em Z}}{\hbox{\cmss Z\kern-.4em Z}}
{\lower.9pt\hbox{\cmsss Z\kern-.4em Z}}
{\lower1.2pt\hbox{\cmsss Z\kern-.4em Z}}\else{\cmss Z\kern-.4em Z}\fi}


\def\IZ{\relax\ifmmode\mathchoice
{\hbox{\cmss Z\kern-.4em Z}}{\hbox{\cmss Z\kern-.4em Z}}
{\lower.9pt\hbox{\cmsss Z\kern-.4em Z}}
{\lower1.2pt\hbox{\cmsss Z\kern-.4em Z}}\else{\cmss Z\kern-.4em
Z}\fi}
\def\IB{\relax{\rm I\kern-.18em B}}
\def\IC{{\relax\hbox{$\inbar\kern-.3em{\rm C}$}}}
\def\ID{\relax{\rm I\kern-.18em D}}
\def\IE{\relax{\rm I\kern-.18em E}}
\def\IF{\relax{\rm I\kern-.18em F}}
\def\IG{\relax\hbox{$\inbar\kern-.3em{\rm G}$}}
\def\IGa{\relax\hbox{${\rm I}\kern-.18em\Gamma$}}
\def\IH{\relax{\rm I\kern-.18em H}}
\def\II{\relax{\rm I\kern-.18em I}}
\def\IK{\relax{\rm I\kern-.18em K}}
\def\IP{\relax{\rm I\kern-.18em P}}

\def\inbar{\,\vrule height1.5ex width.4pt depth0pt}

\font\cmss=cmss10 \font\cmsss=cmss10 at 7pt
\def\IR{\relax{\rm I\kern-.18em R}}


\def\boxit#1{\vbox{\hrule\hbox{\vrule\kern8pt
\vbox{\hbox{\kern8pt}\hbox{\vbox{#1}}\hbox{\kern8pt}}
\kern8pt\vrule}\hrule}}
\def\mathboxit#1{\vbox{\hrule\hbox{\vrule\kern8pt\vbox{\kern8pt
\hbox{$\displaystyle #1$}\kern8pt}\kern8pt\vrule}\hrule}}


\def\inbar{\,\vrule height1.5ex width.4pt depth0pt}

\font\cmss=cmss10 \font\cmsss=cmss10 at 7pt
\def\IR{\relax{\rm I\kern-.18em R}}


\def\a1{{\cal A}^{1,1}}

%

\lref\AFGNT{I. Antoniadis, S. Ferrara, E. Gava, K.S. Narain and
T.R. Taylor, ``Perturbative Prepotential and Monodromies in N=2
Heterotic Superstring'',
{\it Nucl. Phys.} {\bf B447} (1995) 35, {\tt hep-th/9504034}.}

\lref\afgnti{
I. Antoniadis, S. Ferrara, E. Gava, K.S. Narain and
T.R. Taylor,
``Duality Symmetries in N=2 Heterotic Superstring'',
{\tt hep-th/9510079}.}

\lref\aspinwall{
P. S. Aspinwall and J. Louis,
``On the Ubiquity of K3 Fibrations in String Duality'',
{\tt hep-th/9510234}.}

\lref\Candelas{P. Candelas, X. De la Ossa, A. Font, S. Katz and D. Morrison,
``Mirror Symmetry for Two Parameter Models - I'', {\it Nucl. Phys.}
{\bf B416} (1994) 481, {\tt hep-th/9308083}.}

\lref\Cardoso{G. L. Cardoso,  G. Curio,  D. L\"ust,  T. Mohaupt  and  S.-J.
Rey,
``BPS Spectra and Non-Perturbative Gravitational Couplings in N=2, 4
Supersymmetric String Theories'',
{\tt hep-th/9512129}.}

\lref\fhsv{S. Ferrara,  J. A. Harvey,  A. Strominger and  C. Vafa,
``Second-Quantized Mirror Symmetry'', {\it Phys. Lett.} {\bf B361} (1995) 59,
{\tt hep-th/9505162}.}

\lref\gn{V. A. Gritsenko  and  V. V. Nikulin,
``K3 surfaces, Lorentzian Kac--Moody algebras and Mirror Symmetry'',
{\tt alg-geom/9510008}.}

\lref\jt{J. Jorgenson and A. Todorov, ``Analytic Discriminants for 

Manifolds with Canonical Class Zero'', Yale preprint \semi
J. Jorgenson and A. Todorov, ``Enriques surfaces, Analytic discriminants, 

and Borcherds'  $\Phi$ function'', Yale preprint \semi
J. Jorgenson and A. Todorov, ``A Conjectured Analogoue of Dedekind's 

Eta Function for K3 Surfaces'', Math. Res. Lett. {\bf 2} (1995) 359.}

\lref\hm{ J. A. Harvey  and G. Moore,
``Algebras, BPS States, and Strings'',
{\tt hep-th/9510182}.}

\lref\ly{B. H. Lian and S.-T. Yau, ``Mirror Maps, Modular Relations and 

Hypergeometric Series. 2'', {\tt hep-th/9507153}.}

\lref\KLM{A. Klemm, W. Lerche, and P. Mayr, ``K3-Fibrations and
Heterotic-Type II String Duality'', {\it Phys. Lett.} {\bf B357} (1995) 313,
{\tt hep-th/9506112}.}

\lref\KV{S. Kachru and C. Vafa,
``Exact Results for N=2 Compactifications of Heterotic Strings'',
{\it Nucl. Phys.} {\bf B450} (1995) 69, {\tt hep-th/9505105}.}

\lref\KKLMV{
S. Kachru, A. Klemm, W. Lerche, P. Mayr and C. Vafa,
``Nonperturbative Results on the Point Particle
 Limit of N=2 Heterotic String Compactifications'',
{\tt hep-th/9508155}.
}

\lref\kawai{T. Kawai,
``$N=2$ heterotic string threshold correction, $K3$
surface and generalized Kac-Moody superalgebra'',
{\tt hep-th/9512046}.
}

\lref\KLT{ V. Kaplunovsky,  J. Louis and  S. Theisen,
``Aspects of Duality in N=2 String Vacua'',
{\it Phys. Lett.} {\bf B357} (1995) 71, {\tt hep-th/9506110}.
}

\lref\dWKLL{B. de Wit, V. Kaplunovsky, J. Louis and D. L\"ust,
``Perturbative Couplings of Vector Multiplets in N=2 Heterotic String Vacua'',
{\it Nucl. Phys.} {\bf B451} (1995) 53,
{\tt hep-th/9504006}.}
\lref\Yau{S. Hosono, A. Klemm, S. Theisen and S.T. Yau, ``Mirror Symmetry,
Mirror Map and Applications to Calabi-Yau Hypersurfaces'',
{\it Comm. Math. Phys.} {\bf 167} (1995) 301, {\tt hep-th/9308122} \semi
S. Hosono, A. Klemm, S. Theisen and S.T. Yau,
``Mirror Symmetry, Mirror Map, and Applications to Complete Intersection
Calabi-Yau Spaces'', {\it Nucl. Phys.} {\bf B433} (1995) 501, {\tt
hep-th/9406055}.}

\Title{ \vbox{\baselineskip12pt\hbox{{\tt hep-th/9602154}}
\hbox{YCTP-P4/96  }
\hbox{RU-96-10  }}}
{\vbox{
\centerline{Counting Curves   with   Modular Forms }}}
\bigskip
\bigskip
\centerline{M{\aa}ns Henningson and Gregory Moore\footnote{*}{Currently
visiting Rutgers Dept. of Physics.}}

\vskip 1cm
\centerline{\it  Department of Physics}
\centerline{\it Yale University}
\centerline{\it P. O. Box 208120}
\centerline{\it New Haven, CT  06520-8120, USA}
\vskip 3mm
\centerline{mans@genesis5.physics.yale.edu}
\centerline{ moore@castalia.physics.yale.edu}

\bigskip
\bigskip

\centerline{\bf Abstract}
We consider the type IIA string compactified on the Calabi-Yau space given by a
degree 12 hypersurface in the weighted projective space ${\bf P}^4_{(1, 1, 2,
2, 6)}$.
We express the prepotential  of the low-energy effective supergravity theory in
terms of a set of functions that transform covariantly under $PSL(2, \IZ)$
modular transformations. These functions are then determined by monodromy
properties, by
 singularities at the massless monopole point of the moduli space, and by
 $S \leftrightarrow T$ exchange symmetry.

\Date{February 26, 1996; revised April 30, 1996.}

\newsec{Introduction}

Heterotic/Type II  string duality has focused
attention on the special K\"ahler geometry
of vectormultiplets as a means of defining
some nonperturbative effects in the heterotic
string \fhsv\KV.

In this note we describe one
means for obtaining the prepotential for IIA
vectormultiplet geometry using properties
of modular forms.
In particular we focus on the model
described in \KV\  based on IIA
compactification on a manifold
$X(1,1,2,2,6)$ with
 $(h^{1,1}, h^{2,1}) = (2,128)$.
In this example
the K\"ahler cone has two
coordinates $S,T$ and   the special
K\"ahler coordinate  $S$ can be
 identified
with
the heterotic dilaton \KV\aspinwall.   Therefore, on the
heterotic side, nonrenormalization
theorems show that
the (inhomogeneous) vectormultiplet
prepotential has the general form:
\eqn\gnfrm{
{\cal F} (S,T) = ST^2 + f_0(T) + \sum_{k=1}^\infty f_k(T) e^{2\pi i k S}
}
while on  the type IIA side we have:
\eqn\crvecont{
{\cal F} (S,T) = S T^2 + r(T) + {1 \over  (2\pi i )^3} \sum_{j,k}
n_{j,k} Li_3( e^{2 \pi i (j T + k S) } )
}
where $Li_3$ is the trilogarithm,  $n_{j,k}$ counts
rational curves in $X(1,1,2,2,6)$ and
$r(T)$ is a cubic polynomial.

The  prepotential \crvecont\ in this
model has been computed in \Yau\Candelas\
using mirror symmetry.  In this paper we
suggest another method - based on monodromy
and singularity structure - by which one can
determine the prepotential. The procedure may
be viewed as a generalization of the
method
used in  \dWKLL\KLT\afgnti\
to
determine the one-loop prepotential $f_0(T)$
for this model.

In brief, we use the nonperturbative
monodromy  of the special
K\"ahler periods determined in \KKLMV\
to find a set of transformation laws for
the prepotential. The monodromy group
of \KKLMV\  is a discrete subgroup of
$Sp(6;\IZ)$. It acts on the K\"ahler cone, and,
in the limit  $q_2=e^{2 \pi i S} \rightarrow 0$
the action reduces to the standard M\"obius
action of $PSL(2,\IZ)$ on $T$. Therefore,
one may expect the functions $f_k(T)$ to be
related to modular forms for $PSL(2,\IZ)$.
We find that this is indeed so. More precisely,
we can make an upper triangular
transformation of differential polynomials
from $f_k(T)$ to a new set of functions
$h_k(T)$ which are modular forms.
This transformation is summarized in equations
$(3.1)-(3.4)$ below. The relation of the prepotential to 

modular forms has also been investigated in \ly.

Assuming the singularity structure at
$T=i$ implied by the connection
\KV\KKLMV\  to the Seiberg-Witten
massless monopole singularity
we find that $h_k(T)$ can be written
in terms of polynomials of Eisenstein
series.
Finally, using the $T \leftrightarrow S$
symmetry implied by $n_{j,k}=n_{j,j-k}$
\Candelas\   we find that the polynomial
in Eisenstein series is uniquely
determined.  The $T \leftrightarrow S$
symmetry has been the subject of
much recent discussion \KLM\Cardoso.

We must emphasize that the crucial
``upper triangular'' transformation
was discovered ``experimentally'' using
a computer and we have not proved it
analytically, although it has been checked
extensively.

We hope that this work might offer an
alternative method to the standard mirror 

symmetry techniques for counting rational 

curves in (some) Calabi-Yau manifolds, 

which might be of interest in multiparameter
examples. A problem with this approach, 

though, is that one would first have to 

determine the monodromies. 

More speculatively, our methods should help to 

determine root supermultiplicities of  generalized 

Kac-Moody superalgebras \gn\jt\hm\kawai.

\newsec{The monodromy action}

We will consider the type IIA string compactified on the Calabi-Yau threefold
given by a degree $12$ hypersurface in the weighted projective space ${\bf
P}^4_{(1, 1, 2, 2, 6)}$
discussed in \KV\Yau\Candelas.
The degrees of freedom of the low-energy supergravity theory are described by
three vector superfields $X^0$, $X^1$ and $X^2$, corresponding to the
graviphoton and two abelian Yang-Mills multiplets respectively. Their dynamics
are governed by a holomorphic prepotential $F (X^0, X^1, X^2)$. To get the
correct number of propagating fields, $F$ must be homogeneous of degree two in
the $X^i$ \ref\dWvP{
B. de Wit and A. van Proeyen, {\it Nucl. Phys.} {\bf B245} (1984) 89.
}. It is convenient to introduce the inhomogeneous special coordinates $S$ and
$T$, which are defined in terms of the homogeneous coordinates $X^0$, $X^1$ and
$X^2$ as $S = X^1 / X^0$ and $T = X^2 / X^0$. The prepotential can then be
written as
\eqn\prepotential{
F(X^0, X^1, X^2) = (X^0)^2 \left( S T^2 + f (S, T) \right) ,
}
where the first term arises at tree-level and the second term encodes all
(perturbative and non-perturbative) quantum corrections.

We define the periods $F_i$ for $i = 0, 1, 2$ as $F_i = {\partial \over
\partial X^i} F$, i.e.
\eqn\Fzeroonetwo{
\eqalign{
F_0 & = X^0 \left( -S T^2 + 2 f - S {\partial f \over \partial S} - T {\partial
f \over \partial T} \right) \cr
F_1 & = X^0 \left( T^2 + {\partial f \over \partial S} \right) \cr
F_2 & = X^0 \left( 2 S T + {\partial f \over \partial T} \right) , \cr
}
}
and assemble the homogeneous coordinates and the periods in a period vector
$\Pi$ given by
$\Pi = \left( X^0 \;\;\; X^1 \;\;\; X^2 \;\;\; F_0 \;\;\; F_1 \;\;\; F_2
\right)^T$. As we encircle a singular divisor in the moduli space, the period
vector $\Pi$ is acted on by multiplication by an element of the monodromy
group. This group is a subgroup of $Sp (6, {\bf Z})$ generated by three
elements $S_1$, $T_1$ and $T_2$ \KKLMV , which in our basis are given by
\eqn\monodromy{
\eqalign{
S_1 & = \left( \matrix{
0 & 0 & 0 & 0 & -1 & 0 \cr -1 & 0 & 0 & 1 & -1 & 0 \cr 0 & 0 & 1 & 0 & 0 & 0
\cr 1 & 1 & 0 & 0 & 1 & 0 \cr -1 & 0 & 0 & 0 & 0 & 0 \cr 0 & 0 & 0 & 0 & 0 & 1
\cr
} \right) \cr
T_1 & = \left( \matrix{
1 & 0 & 0 & 0 & 0 & 0 \cr 0 & 1 & 0 & 0 & 0 & 0 \cr 1 & 0 & 1 & 0 & 0 & 0 \cr
-5 & -1 & -4 & 1 & 0 & -1 \cr 1 & 0 & 2 & 0 & 1 & 0 \cr 0 & 2 & 4 & 0 & 0 & 1
\cr
} \right) \cr
T_2 & = \left( \matrix{
1 & 0 & 0 & 0 & 0 & 0 \cr 1 & 1 & 0 & 0 & 0 & 0 \cr 0 & 0 & 1 & 0 & 0 & 0 \cr 0
& 0 & 0 & 1 & -1 & 0 \cr 0 & 0 & 0 & 0 & 1 & 0 \cr 0 & 0 & 2 & 0 & 0 & 1 \cr
} \right) . \cr
}
}
The monodromies under $T_1,T_2$ can be deduced, on the
type II side, from the monodromy of the mirror manifold at
$\infty$ in complex structure space. The monodromy under
$S_1$ must then be deduced, on the type II side, by
solving Picard-Fuchs equations. Alternatively,  assuming
string/string duality, it may be deduced, on the
heterotic side,  from the one-loop
approximation to $f(S,T)$.

We will now determine the action of the monodromy group on the special
coordinates $S$ and $T$ and the function $f$ appearing in the prepotential
\prepotential. It follows from the above that the $T_2$ transformation acts as
\eqn\Ttwo{
\eqalign{
S & \rightarrow S + 1 \cr
T & \rightarrow T \cr
f & \rightarrow f  . \cr
}
}
This means that $f$ may be expanded in powers of $q_2 = \exp 2 \pi i S$
\ref\CDAF{
A. Ceresole, R. D'Auria and S. Ferrara, {\it Phys. Lett.} {\bf B339} (1994) 71,
{\tt hep-th/9408036}.
}, i.e.
\eqn\fexpansion{
f (S, T) = \sum_{k = 0}^\infty q_2^k f_k (T)
}
for some functions $f_k$. The $T_1$ transformation acts as
\eqn\Tone{
\eqalign{
S & \rightarrow S \cr
T & \rightarrow T + 1 \cr
f  & \rightarrow f + 2 T^2 - {5 \over 2} . \cr
}
}
In terms of the functions $f_k$ in \fexpansion, this means that
\eqn\Tshiftzero{
f_0 (T + 1)  = f_0 (T) + 2 T^2 - {5 \over 2}
}
and
\eqn\Tshiftk{
f_k (T + 1)  = f_k (T) , \;\;\;\; k > 0 .
}

The consequences of the $S_1$ transformation are less straightforward to
extract. It acts as
\eqn\Sone{
\eqalign{
S & \rightarrow \tilde{S} = 1 + {S + T^{-2} - 2 T^{-2} f + S T^{-2} {\partial f
\over \partial S} + T^{-1} {\partial f \over \partial T} \over 1 +T^{-2}
{\partial f \over \partial S}} \cr
T & \rightarrow \tilde{T} = - {T^{-1} \over 1 + T^{-2} {\partial f \over
\partial S}} \cr
}
}
and
\eqn\ftransf{
f \rightarrow {T^{-4} f - {1 \over 2} T^{-4}  - T^{-2}  - {1 \over 2} + \ldots
\over \left(1 + T^{-2} {\partial f \over \partial S} \right)^3} ,
}
where the omitted terms are proportional to powers of ${\partial f \over
\partial S}$. Inserting \fexpansion\ and \Sone\ in \ftransf\ and taking the $S
\rightarrow i \infty$ limit, we find that
\eqn\fzerotransf{
f_0 (- 1 / T) = T^{-4} f_0 (T) - {1 \over 2} T^{-4} - T^{-2} - {1 \over 2} .
}
The transformation properties of the functions $f_k$ for $k > 0$ are most
easily deduced by considering ${\partial f \over \partial S}$, which transforms
under the $S_1$ transformation as
\eqn\fstransf{
{\partial f \over \partial S} \rightarrow {T^{-4} {\partial f \over \partial S}
\over \left(1 + T^{-2} {\partial f \over \partial S} \right)^2} .
}
Inserting \fexpansion\ and \Sone\ in \fstransf, we get
\eqn\sumtransf{
\sum_{k = 1}^\infty 2 \pi i k f_k (\tilde{T}) \exp 2 \pi i k \tilde{S} =
{T^{-4} \sum_{k = 1}^\infty 2 \pi i k f_k (T) \exp 2 \pi i S \over \left( 1 +
T^{-2} \sum_{k = 1}^\infty 2 \pi i k f_k (T) \exp 2 \pi i S \right)^2} ,
}
where $\tilde{S}$ and $\tilde{T}$ are given in \Sone. By expanding this
equation in powers of $q_2 = \exp 2 \pi i S$ and expanding the functions $f_k$
on the left hand side in a Taylor series around $- 1 / T$, we can recursively
determine the transformation laws of these functions. For example, for $f_1$
and $f_2$ we get
\eqn\fonetwo{
\eqalign{
f_1 (- 1 / T) & = \exp 2 \pi i \left(-T^{-2} + 2 T^{-2} f_0 (T) - T^{-1}
f_0^{(1)} (T) \right) T^{-4} f_1 (T) \cr
f_2 (- 1 / T) & = \exp 4 \pi i \left(-T^{-2} + 2 T^{-2} f_0 (T) - T^{-1}
f_0^{(1)} (T) \right) \cr
& \;\;\;\;\;\;\;\; \times \Bigl(T^{-4}  f_2 (T) \cr
& \;\;\;\;\;\;\;\;\;\;\;\;\;\; + T^{-5} (- 2 \pi i f_1 (T) f_1^{(1)} (T)) \cr
& \;\;\;\;\;\;\;\;\;\;\;\;\;\; + T^{-6} (4 \pi i f_1 (T) f_1 (T) - 2 \pi^2 f_1
(T) f_1 (T) f_0^{(2)} (T)) \cr
& \;\;\;\;\;\;\;\;\;\;\;\;\;\; + T^{-7} 4 \pi^2 f_1 (T) f_1 (T) f_0^{(1)} (T)
\cr
& \;\;\;\;\;\;\;\;\;\;\;\;\;\; + T^{-8} (2 \pi^2 f_1 (T) f_1 (T) - 4 \pi^2 f_0
(T) f_1 (T) f_1 (T)) \Bigr)  \cr
\ldots \cr
}
}
where superscripts in parenthesis indicate derivatives with respect to $T$.

\newsec{Transformation to modular forms}

The transformation laws \Tshiftzero, \Tshiftk, \fzerotransf\ and \fonetwo\
indicate that the functions $f_k$ are closely related to modular forms. Indeed,
it follows from \Tshiftzero\ and \fzerotransf\ that $f_0^{(5)}$ is a modular
form of weight $6$ \dWKLL\AFGNT.

To see how the $f_k$ for $k > 0$ are related to modular forms, we first
introduce a new formal expansion parameter $q$ (  distinct from $q_1 = \exp 2
\pi i T$ and $q_2 = \exp 2 \pi i S$) and a function $h$ with an expansion of
the form
\eqn\h{
h = \sum_{k = 0}^\infty q^k h_k (T) .
}
Next, we define a set of functions $g_{(m)}$ for $m = 0, 1, \ldots$ recursively
by
\eqn\g{
\eqalign{
g_{(0)}  & = {\partial \over \partial S} f \cr
g_{(m + 1)} & = \sum_{n = 0}^\infty {1 \over n!} {1 \over m n + 1} \left(
{(2m-1) (-1)^m \over 6 (2m)!} h^{(2 m + 2)} \right)^n {\partial^n \over
\partial S^n} ( g_{(m)})^{m n + 1} , \cr
}
}
where again superscripts in parenthesis denote derivatives with respect to $T$.
Finally, we {\it define} the relationship between the functions $h$ and $f$ to
be given by the equation
\eqn\hf{
h = \lim_{m \rightarrow \infty} \left. \left( {\partial \over \partial S}
\right)^{-1} g_{(m)} \right|_{q_2 = q} ,
}
where the integration constant arising from the $\left( {\partial \over
\partial S} \right)^{-1}$ operator is fixed by the requirement that the $q$
independent term of the right-hand side of \hf\ equal $f_0 (T)$.
Our  conjecture is now that
the $h_k(T)$ for $k > 0$ transform covariantly with weight $-4$ under
the modular group:
\eqn\htransf{
\eqalign{
h_k (T + 1) & = \exp \left( {2 \pi i k \over 3} \right) h_k (T) \cr
h_k (- 1 / T) & = \exp \left( {2 \pi i k \over 3} \right) T^{-4} h_k (T) . \cr
}
}
We have no analytic proof of this statement, but it has
been checked by computer up to $k=7$.

The explicit formulae for the $h_k(T)$ can be obtained as follows.
We begin by expanding \g\ to get:
\eqn\gzeroonetwo{
\eqalign{
g_{(0)} & = \sum_{k = 1}^\infty q_2^k 2 \pi i k f_k \cr
g_{(1)} & = \sum_{k = 1}^\infty q_2^k \exp \left( - { \pi i k \over 3} h^{(2)}
\right) 2 \pi i k f_k \cr
g_{(2)} & = \sum_{k = 1}^\infty q_2^k \exp \left( - { \pi i k \over 3} h^{(2)}
\right) \sum_{n = 0}^\infty  {1 \over (n + 1)!} \left( - {\pi i k \over 6}
h^{(4)} \right)^n \cr
& \;\;\;\;\;\;\;\;\;\;\;\; \times \sum_{k_0, \ldots, k_n = 1}^\infty
\delta_{k_0 + \ldots + k_n, k} \prod_{i = 0}^n 2 \pi i k_i f_{k_i} \cr
\ldots
}
}
Then,  inserting \h\ in \hf\ and solving recursively
for the $f_k$ in terms of the $h_k$, we   get
\eqn\fh{
\eqalign{
f_0 & = h_0 \cr
f_1 & = \exp \left( {\pi i \over 3} h_0^{(2)} \right) h_1 \cr
f_2 & = \exp \left( {2 \pi i \over 3} h_0^{(2)} \right) \left(h_2 + {\pi i
\over 3} h_1 h_1^{(2)} - {\pi^2 \over 6} h_1 h_1 h_0^{(4)} \right) \cr
f_3 & = \exp \left( {3 \pi i \over 3} h_0^{(2)} \right) \left( h_3 + {2 \pi i
\over 3} h_2 h_1^{(2)} - {\pi^2 \over 6} h_1 h_1^{(2)} h_1^{(2)} + {\pi i \over
3} h_1 h_2^{(2)} \right. \cr
& \;\;\;\;\;\;\;\;\;\;\;\;\;\;\; \left. - {2 \pi^2 \over 3} h_1 h_2 h_0^{(4)}
- {2 \pi^3 i \over 9} h_1 h_1 h_1^{(2)} h_0^{(4)} + {\pi^4 \over 18} h_1 h_1
h_1 h_0^{(4)} h_0^{(4)} \right. \cr
& \;\;\;\;\;\;\;\;\;\;\;\;\;\;\; \left. - {\pi^2 \over 6} h_1 h_1 h_1^{(4)} +
{\pi^3 i \over 18} h_1 h_1 h_1 h_0^{(6)} \right)
}
}
$$
\eqalignno{
f_4 & = \exp \left( {4 \pi i \over 3} h_0^{(2)} \right) \left( h_4 + \pi i h_3
h_1^{(2)} - {4 \pi^2 \over 9} h_2 h_1^{(2)} h_1^{(2)} - {8 \pi^3 i \over 81}
h_1 h_1^{(2)} h_1^{(2)} h_1^{(2)} \right. \cr
& \;\;\;\;\;\;\;\;\;\;\;\;\;\;\; \left. + {2 \pi i \over 3} h_2 h_2^{(2)} - {4
\pi^2 \over 9} h_1 h_1^{(2)} h_2^{(2)} + {\pi i \over 3} h_1 h_3^{(2)} - {2
\pi^2 \over 3} h_2 h_2 h_0^{(4)} \right. \cr
& \;\;\;\;\;\;\;\;\;\;\;\;\;\;\; \left. - \pi^2 h_1 h_3 h_0^{(4)} - {10 \pi^3 i
\over 9} h_1 h_2 h_1^{(2)} h_0^{(4)} + {13 \pi^4 \over 54} h_1 h_1 h_1^{(2)}
h_1^{(2)} h_0^{(4)} \right. \cr
& \;\;\;\;\;\;\;\;\;\;\;\;\;\;\; \left. - {\pi^3 i \over 3} h_1 h_1 h_2^{(2)}
h_0^{(4)} + {4 \pi^4 \over 9} h_1 h_1 h_2 h_0^{(4)} h_0^{(4)} \right. \cr
& \;\;\;\;\;\;\;\;\;\;\;\;\;\;\; \left. + {4 \pi^5 i \over 27} h_1 h_1 h_1
h_1^{(2)} h_0^{(4)} h_0^{(4)} - {2 \pi^6 \over 81} h_1 h_1 h_1 h_1 h_0^{(4)}
h_0^{(4)} h_0^{(4)} \right. \cr
& \;\;\;\;\;\;\;\;\;\;\;\;\;\;\; \left. - {2 \pi^2 \over 3} h_1 h_2 h_1^{(4)} -
{5 \pi^3 i \over 18} h_1 h_1 h_1^{(2)} h_1^{(4)} + {\pi^4 \over 6} h_1 h_1 h_1
h_0^{(4)} h_1^{(4)} \right. \cr
& \;\;\;\;\;\;\;\;\;\;\;\;\;\;\; \left. - {\pi^2 \over 6} h_1 h_1 h_2^{(4)} +
{\pi^3 i \over 3} h_1 h_1 h_2 h_0^{(6)} - {\pi^4 \over 9} h_1 h_1 h_1 h_1^{(2)}
h_0^{(6)} \right. \cr
& \;\;\;\;\;\;\;\;\;\;\;\;\;\;\; \left. - {\pi^5 i \over 18} h_1 h_1 h_1 h_1
h_0^{(4)} h_0^{(6)} + {\pi^3 i \over 18} h_1 h_1 h_1 h_1^{(6)} + {\pi^4 \over
216} h_1 h_1 h_1 h_1 h_0^{(8)} \right) \cr
\ldots
}
$$

(To determine $f_k$ for a given $k$, it is in fact enough to let $m = k$ in
\hf\ rather than taking the limit $m \rightarrow \infty$.) Assigning a formal
weight $-4 + 2 n$ and a charge $k$ to $h_k^{(n)}$, we see that the general
structure is that $f_k$ equals $\exp \left( {k \pi i \over 3} h_0^{(2)}
\right)$ times a differential polynomial in $h$ of formal weight $-4$ and
charge $k$ involving only even derivatives. The `upper triangular' structure of
these equations makes them easy to invert:
\eqn\hk{
\eqalign{
h_0 & = f_0 \cr
h_1 & = \exp \left( - {\pi i  \over 3} f_0^{(2)} \right) f_1 \cr
h_2 & = \exp \left( - {2 \pi i  \over 3} f_0^{(2)} \right) \cr
& \;\;\;\;\;\; \times \left( f_2 - {\pi i \over 3} f_1^{(2)} f_1 - {2 \pi^2
\over 9} f_0^{(3)} f_1^{(1)} f_1 + {\pi^2 \over 18} f_0^{(4)} f_1 f_1 + {\pi^3
i \over 27} f_0^{(3)} f_0^{(3)} f_1 f_1 \right) \cr
\ldots
}
}

\newsec{The singularity structure}

Next, we need some facts about the singularities of the functions $f_k$. These
occur at $T = i$, where the gauge group is enlarged and additional multiplets
become massless. As $T \rightarrow i$, $f_0^{(2)} (T)$ diverges as $-{8 \over 2
\pi i} \log (T - i)$ \dWKLL \AFGNT \ref\CLM{
G. L. Cardoso, D. L\"ust and T. Mohaupt, ``Threshold Corrections and Symmetry
Enhancement in String Compactifications'', {\it Nucl. Phys.} {\bf B450} (1995)
115, {\tt hep-th/9412209}.
}. Furthermore, to recover the results of \ref\SW{
N. Seiberg and E. Witten, ``Electric-Magnetic Duality, Monopole Condensation,
and Confinement in $N = 2$ Supersymmetric Yang-Mills Theory'', {\it Nucl.
Phys.} {\bf B426} (1994) 19, {\tt hep-th/9407087}.
} in the limit when the string tension becomes large, $f_k (T)$ for $k > 0$
must have a pole of order $4 k - 2$ at $T = i$ \KKLMV. Finally, it follows from
\Tshiftzero\ that $f_0^{(2)}$ diverges as $4 T$ as $T \rightarrow i \infty$.
{}From the general form of the functions $h_k$ \hk, we then conclude that their
singularities are given by
\eqn\hsingularities{
\eqalign{
h_k (T) & \sim (T - i)^{2 - 8 k / 3} , \;\;\;\;\;\;  T \rightarrow i \cr
h_k (T) & \sim q_1^{- 2 k / 3}, \;\;\;\;\;\;  T \rightarrow i \infty . \cr
}
}

{}From the fact that $f_0^{(5)}$ is a modular form of weight $6$ and the above
singularity structure, it can be determined  \KLT\afgnti\ as
\eqn\fzerofive{
f_0^{(5)} = (2 \pi i)^2 {18 E_4^6 - 23 E_4^3 E_6^2 + 5 E_6^4 \over 9 E_6^3} ,
}
where the Eisenstein series of weight $4$ and $6$ are defined as
\eqn\Eisenstein{
\eqalign{
E_4 (T) & = 1 + 240 \sum_{j = 1}^\infty {j^3 q_1^j \over 1 - q_1^j} \cr
E_6 (T) & = 1  - 504 \sum_{j = 1}^\infty {j^5 q_1^j \over 1 - q_1^j} \cr
}
}
for $q_1 = \exp 2 \pi i T$. They have simple zeros at $T = \exp (\pi i / 3)$
and at $T = i$ respectively. Integrating $f_0^{(5)}$ five times gives $f_0$;
all integration constants except one are determined by imposing \Tshiftzero:
\eqn\fzero{
f_0 (T) = {\rm constant} - {13 \over 6} T - T^2  + {2 \over 3} T^3  + {\cal O}
(q_1) .
}

{}From \htransf\ and \hsingularities, we conclude that for $k > 0$
\eqn\hstructure{
h_k = (2 \pi i)^{-3} {P_{24 k - 16} \over \eta^{16 k} E_6^{8 k / 3 - 2}} ,
}
where the $\eta$ invariant is defined as
\eqn\etadef{
\eta (T) = q_1^{1 / 24} \prod_{j = 1}^\infty \left( 1 - q_1^j \right)
}
and $P_{24 k - 16}$ is a holomorphic modular form of weight $24 k - 16$. As
such, $P_{24 k - 16}$ must be of the form
\eqn\Pstructure{
P_{24 k - 16} = \sum_{n = 1}^{2 k - 1} p_{k, n} E_4^{3 n - 1} E_6^{4 k - 2 n -
2} ,
}
for some constants $p_{k, 1}, \ldots, p_{k, 2 k - 1}$.

\newsec{The exchange symmetry}

To determine the forms $P_{24 k - 16}$ in \hstructure, we consider the Yukawa
coupling $\left( {\partial \over \partial T} \right)^3 f$, which may be written
in the form \ref\AM{
P. S. Aspinwall and D. R. Morrison, ``Topological Field Theory and Rational
Curves'', {\it Comm. Math. Phys.} {\bf 151} (1993) 245, {\tt hep-th/9110048}.
}
\eqn\Yukawa{
{\partial^3 \over  \partial T^3} f = 4 + \sum_{k = 0}^\infty \sum_{j =
1}^\infty j^3 n_{j, k} {q_1^j q_2^k \over 1 - q_1^j q_2^k} .
}
For fixed $k > 0$, the $S \leftrightarrow T$ exchange symmetry
\Candelas\KLM\Cardoso\ amounts to $2 k - 1$ constraints on the instantonnumbers
$n_{j, k}$ in \Yukawa:
\eqn\constraints{
n_{j, k}  = \left\{ \matrix{ 0 & , & 1 \leq j \leq k - 1 \cr n_{j, j - k} & , &
k \leq j \leq 2 k - 1 \cr } \right. .
}
The $h_k$ can now be determined recursively: Given $h_1, \ldots, h_{k - 1}$,
the constraints \constraints\ determine the constants $p_{k, 1}, \ldots, p_{k,
2 k - 1}$ in \Pstructure\ and thus, by \hstructure, $h_k$. Explicitly, the
first few $h_k$'s are given by
\eqn\hvalue{
\eqalign{
h_1 & = (2 \pi i)^{-3} \eta^{-16} E_6^{-2 / 3} 2 E_4^2 \cr
h_2 & = (2 \pi i)^{-3} 2^{-4} 3^{-3} \eta^{-32} E_6^{-10 / 3} \left(-89 E_4^8 -
53 E_4^5 E_6^2  + 122 E_4^2 E_6^4 \right) \cr
h_3 & = (2 \pi i)^{-3} 2^{-5} 3^{-7} \eta^{-48} E_6^{-6} \cr
& \;\;\;\;\;\;\; \times \left( 20367 E_4^{14} - 38052 E_4^{11} E_6^2 + 18898
E_4^8 E_6^4 - 6260 E_4^5 E_6^6 + 3895 E_4^2 E_6^8 \right) \cr
h_4 & = (2 \pi i)^{-3} 2^{-13} 3^{-10} \eta^{-64} E_6^{-26 / 3} \cr
& \;\;\;\;\;\;\; \times \left( 216412213 E_4^{20} - 793763223 E_4^{17} E_6^2 +
1110594390 E_4^{14} E_6^4 \right. \cr
& \;\;\;\;\;\;\;\;\;\;\;\;\; \left. - 711685317 E_4^{11} E_6^6 + 217366407
E_4^8 E_6^8 - 18944802 E_4^5 E_6^{10} \right. \cr
& \;\;\;\;\;\;\;\;\;\;\;\;\; \left. + 5991276 E_4^2 E_6^{12} \right) \cr
\ldots
}
}
(We have also calculated $h_5$, $h_6$ and $h_7$.) Inserting this result in \fh\
gives the functions $f_k$ in the prepotential. The instanton expansion \Yukawa\
then gives the instantonnumbers $n_{j, k}$, which count rational curves in the
Calabi-Yau manifold:
$$
\halign{\hfil # & # & \hfil # & \hfil # & \hfil # & \hfil # \cr
j & & k = 0 & k = 1 & k = 2 &      . . .  \cr
\cr
0 & & 0 & 2 & 0 \cr
1 & & 2496 & 2496 & 0 \cr
2 & & 223752 & 1941264 & 223752 \cr
3 & & 38637504 & 1327392512 & 1327392512 \cr
4 & & 9100224984 & 861202986072 & 2859010142112 \cr
5 & & 2557481027520 & 540194037151104 &  4247105405354496 \cr
6 & & 805628041231176 & 331025557765003648 & 5143228729806654496 \cr
7 &  & 274856132550917568 & 199399229066445715968 & 5458385566105678112256 \cr
. . . \cr
}
$$
These numbers agree with the results of \Candelas.

Thus, we have established the claim that the monodromy,
singularity structure, and $S\leftrightarrow T$ exchange symmetry
 are sufficient to determine the prepotential exactly in the region where the
series converges.

\bigskip
\bigskip
\bigskip
\centerline{\bf Acknowledgements}
We would like to thank J. Harvey for participation at
the beginning of this project and for discussions. We also thank
 P. Berglund, A. Losev, D. Morrison, N. Nekrasov, R. Plesser,
and S. Shatashvili for
useful correspondence and discussions.
This research
is supported by DOE grant DE-FG02-92ER40704,
and by a Presidential Young Investigator Award.
GM also thanks the Rutgers high energy theory group
for hospitality during the completion of this paper.

\listrefs

\bye